\documentclass[12pt]{article}
\pdfoutput=1
\usepackage{amssymb,graphicx}
\usepackage{epstopdf}
\usepackage{amsmath,amsfonts}
\usepackage{epsfig}
\usepackage{graphicx,graphics}
\usepackage{tikz}
\usetikzlibrary{trees}
\usetikzlibrary{decorations.pathmorphing}
\usetikzlibrary{decorations.markings}

\begin{document}

\vskip 10cm

\sloppy

\title{\bf Living with ghosts in Ho\v{r}ava--Lifshitz gravity}
\author{S.~Ramazanov$^a$\footnote{{\bf e-mails}:
    sabir.ramazanov@gssi.infn.it, arroja@phys.ntu.edu.tw, marco.celoria@gssi.infn.it, sabino.matarrese@pd.infn.it, luigi.pilo@aquila.infn.it}, F.~Arroja$^b$, M.~Celoria$^a$, S.~Matarrese$^{c,a,d, e}$, L.~Pilo$^{f,g}$\\
 \small{$^a$\em Gran Sasso Science Institute (INFN), Viale Francesco Crispi 7, I-67100 L'Aquila, Italy}\\
 \small{$^b$\em Leung Center for Cosmology and Particle Astrophysics, National Taiwan University,}\\
 \small{\em No.~1, Sec.~4, Roosevelt Road, Taipei, 10617 Taipei, Taiwan (R.O.C.)}\\
 \small{$^c$\em Dipartimento di Fisica e Astronomia ''G.~Galilei'',}\\
 \small{\em Universit\'a degli Studi di Padova, via Marzolo 8, I-35131 Padova, Italy}\\
 \small{$^d$\em INFN, Sezione di Padova, via Marzolo 8, I-35131 Padova, Italy}\\
 \small{$^e$\em INAF-Osservatorio Astronomico di Padova, Vicolo dell'Osservatorio 5, I-35122 Padova, Italy}\\
 \small{$^f$\em Dipartimento di Scienze Fisiche e Chimiche, Universit\'a di L'Aquila, I-67010 L'Aquila, Italy}\\
 \small{$^g$ \em INFN, Laboratori Nazionali del Gran Sasso, I-67010 Assergi, Italy}
 }

{\let\newpage\relax\maketitle}

\begin{abstract}
We consider the branch of the projectable Ho\v{r}ava-Lifshitz model which exhibits
ghost instabilities in the low energy limit.
It turns out that,  due to the Lorentz violating structure of the model and to the presence of a
finite strong coupling scale, the vacuum decay rate into  photons is
tiny in a wide range of phenomenologically acceptable parameters. The
strong coupling scale, understood as a cutoff on ghosts' spatial momenta, can be raised up to
$\Lambda \sim 10$ TeV.
At lower momenta, the projectable Ho\v{r}ava-Lifshitz gravity is
equivalent to  General Relativity supplemented by a fluid with a small positive sound speed squared ($10^{-42}\lesssim$) $c^2_s \lesssim 10^{-20}$, that
could be a promising candidate for the Dark Matter. Despite these advantages, the {\it unavoidable} presence of the
strong coupling obscures the implementation of the original Ho\v{r}ava's proposal on
quantum gravity. Apart from the Ho\v{r}ava-Lifshitz model,
conclusions of the present work hold also for the mimetic matter scenario,
where the analogue of the projectability condition is achieved by a non-invertible
conformal transformation of the metric.

\end{abstract}

\section{Introduction and Summary}

Modifications of gravity are the subject of intensive debate in the scientific
community. The main interest comes from cosmology.
In particular, we do not know yet the origin of $95\%$ of the energy density
in the Universe. While the Dark Matter component (27\%) can be relatively easily
explained by an extension of the Standard Model, no convincing
model of Dark Energy (68\%) has been proposed yet~\cite{Ade:2015xua}.
Besides these cosmological considerations, finding a viable
modification of General Relativity (GR) is, in a certain sense, a matter of principle.
The first attempts to promote the graviton to a
massive field (the renowned Fierz--Pauli model,~\cite{Pauli}), have been made well before the
discovery of Dark Matter and Dark Energy. It turned out, however,
that the Fierz--Pauli model is plagued by ghost instabilities~\cite{Deser} and an unacceptably
low scale of the strong coupling~\cite{Vainshtein}. Motivated by those problems,
several proposals on Lorentz-invariant models of gravity have been put forward
recently~\cite{Hinterbichler:2011tt, deRham:2014zqa}.

Another interesting approach to modifying gravity
is to break Lorentz-invariance spontaneously or explicitly. (See the
review~\cite{Rubakov:2008nh}, the references therein and  Refs.~\cite{Comelli:2014xga,Comelli:2013txa,Comelli:2013tja}).
This is argued to be a viable way of getting around the troubles with ghost instabilities and a low scale of the strong coupling~\cite{Rubakov:2004eb, Cline:2003gs, Dubovsky:2004sg}. Moreover,
relaxing the Lorentz symmetry, one opens up an intriguing opportunity to construct Dark Energy, inflation and Dark Matter, with a row of interesting models exemplified in the ghost
condensate context~\cite{ArkaniHamed:2003uy, ArkaniHamed:2003uz, ArkaniHamed:2005gu}.

Violation of Lorentz invariance at the high energies may have
dramatic consequences in a view of another long-standing problem---the renormalization of gravity. Following Ho\v{r}ava's proposal~\cite{Horava:2009uw, Horava:2008ih, Horava:2009if}, one assigns different scaling dimensions to the
time, $t$, and spatial coordinates, $\bf x$, in the ultraviolet (UV),
\begin{equation}
\label{scaling}
{\bf x} \rightarrow b^{-1}{\bf x}\;, \qquad t \rightarrow b^{-z} t \; ,
\end{equation}
where $z$ is the so-called critical exponent. The scaling~\eqref{scaling} fixes the high energy dispersion relation for the
graviton to be of the form $\omega^2 \propto {\bf p}^{2z}$, where $\omega$ denotes the energy and $\bf p$ the momentum. Consequently, the behaviour
of the propagators changes in the UV, potentially eliminating
the divergence of the loop integrals. Power-counting (super-)renormalizable gravity
corresponds to the choice $z=3$ ($z>3$). See the discussion in Subsection 2.1.

However, breaking the Lorentz symmetry does not come at zero price. It implies the
smaller group of diffeormorphisms compared to that of GR. As an immediate
consequence, one has extra degrees of freedom, on top of the
standard helicity-2  graviton. In particular, the lapse function, $N(t, {\bf x})$, is a
dynamical variable in Ho\v{r}ava--Lifshitz gravity. This new degree of freedom
suffers from gradient instabilities
and furthermore leads to a very low strong coupling scale in the theory~\cite{Charmousis:2009tc, Blas:2009yd}. Looking for the solution of the problem, one either turns to extensions of the Ho\v{r}ava--Lifshitz gravity or imposes
some conditions capable of eliminating the pathological mode. The first line of research
was developed in Ref.~\cite{Blas:2009qj}, and we do not touch it in the present paper. We follow
another approach, proposed already in the original paper by Ho\v{r}ava~\cite{Horava:2009uw}, which is to impose the projectability condition.

In the projectable Ho\v{r}ava--Lifshitz gravity one entertains the possibility
that the lapse $N$ is a function of time only, i.e., $N=N(t)$. The reason why this
scenario is important is twofold. First, it is the only version of the
Ho\v{r}ava--Lifshitz gravity which has been renormalized so far~\cite{Barvinsky:2015kil}\footnote{However, in 4 dimensions the perturbative renormalization of the projectable Ho\v rava--Lifshitz gravity is achieved at the price of the phenomenological viability
of the model. See the discussion below. }. Second, it has quite a rich phenomenology. In particular, this model provides a candidate for the Dark Matter and an alternative to inflation~\cite{Mukohyama:2009mz, Mukohyama:2009zs, Mukohyama:2010xz, Mukohyama:2009gg, Mukohyama:2009tp}.

Still, the model of interest appears to be problematic in the infrared limit (IR).
At low spatial momenta, it is equivalent to GR plus a fluid characterized by the constant sound speed $c_s$~\cite{Blas:2009yd, Blas:2010hb, Koyama:2009hc},--- a statement, which becomes
manifest upon performing the Stuckelberg trick~\cite{Blas:2009yd, Blas:2010hb, Germani:2009yt} (see Subsection 2.2).
In what follows, we will consider two branches of the projectable Ho\v{r}ava--Lifshitz
gravity depending on the sign of the quantity $c^2_s$.
The branch with the negative sound speed squared, i.e., $c^2_s<0$, is the only studied in the literature. It is plagued
by gradient instabilities in the IR, which can be cured in the UV by the Lorentz-violating operators inherent in Ho\v{r}ava--Lifshitz
gravity. Naively, by tuning the sound speed $c_s$ to be arbitrarily small,
one could avoid any conflict with observations. This expectation, however,
is not met in reality:
the strong coupling scale is finite and depends crucially on the sound speed~\cite{Blas:2010hb, Koyama:2009hc},
\begin{equation}
\label{str}
\Lambda_p \sim M_{Pl} \cdot |c_s|^{3/2} \; .
\end{equation}
Here $M_{Pl}$ denotes the Planck mass defined as $M_{Pl} \equiv G^{-1/2}$,
where $G$ is Newton's constant. Note that the cutoff~\eqref{str} is applied to the
spatial momenta $p \equiv |{\bf p}|$, i.e., it breaks Lorentz invariance explicitly.  Demanding that gradient instabilities do not propagate on a
time scale smaller than the age of the Universe, leads to the constraint $\Lambda_p \lesssim 10^{-17}$ eV~\cite{Blas:2010hb}. This is by many orders of magnitude below the phenomenologically allowed values.

In this paper, we will argue that the second branch, with $c_s^2>0$, of the Ho\v{r}ava--Lifshitz gravity
is viable in a much wider range of sound speed values, so that
the model remains perturbative down to microscopic scales.
The reason, why this branch has been ignored in the literature is simple:
unlike standard cosmological fluids, e.g., radiation, in our case the positive sound speed squared comes at the price of having ghost instabilities~\cite{Blas:2009qj, Blas:2010hb, Koyama:2009hc}.
Naively, this invalidates the scenario of interest. However, it
is well-known that ghosts are not
particularly dangerous in Lorentz-violating theories of gravity~\cite{Cline:2003gs}\footnote{The opportunity to have controllable ghost instabilities in Lorentz-invariant theories is discussed in Ref.~\cite{Garriga:2012pk}.}. That is, for a reasonably small cutoff $\Lambda$ on the momenta and the frequencies of the ghosts,
the decay of the vacuum into Standard Model species in the final state is
sufficiently slow,---in agreement with the observed abundance of particles in the
Universe\footnote{See Refs.~\cite{Kaplan:2005rr} and~\cite{Dyda:2012rj} for the applications of this idea to various gravitational frameworks.}.  In the projectable Ho\v{r}ava--Lifshitz gravity that cutoff is realized
by the finite scale of the strong coupling~\eqref{str}, above which, we assume, the
model is free of instabilities. The upper limit on the scale $\Lambda$ obtained in Ref.~\cite{Cline:2003gs} reads $\Lambda \lesssim 3~\mbox{MeV}$.
The latter is applied provided that the frequency $\omega$ and the spatial momenta ${\bf p}$ have equal cutoffs. This
condition, however, is not fulfilled in our scenario, where the ghosts obey the non-relativistic
dispersion relation $\omega^2 =c^2_s {\bf p}^2$, with the sound speed much less than unity, i.e.,
$c_s \ll 1$. As a by-product,
the cutoff on the spatial momenta (strong coupling scale~\eqref{str}) can be relaxed by many orders of magnitude, extending the validity of perturbation theory to the TeV range (see Section~3).

The reason is twofold. First, in the formal limit $c_s \rightarrow 0$, the energies of the ghosts and, say, the photons produced from the vacuum decay approach zero. This leads to a vanishing phase-space volume of the
outgoing particles. For generic values of the sound speed (still, much smaller than unity), this argument implies the suppression of the phase-space volume by some power of the quantity $c_s$. Second,
the coupling between the standard fields,
i.e., photons and helicity-2 gravitons, with the canonical ghost field is very weak. This follows from normalization
considerations. Collecting the suppressing factors altogether and comparing the resulting decay rate into photons with the observed
flux of cosmic X-rays, one obtains $c^2_s \lesssim 10^{-20}$. The upper limit here implies that the strong coupling scale in the projectable Ho\v{r}ava--Lifshitz
gravity can be as large as $\Lambda_p \sim 10$ TeV. This is about thirty orders of magnitude weaker than
the associated limit in the branch plagued by the gradient instabilities.

To summarize, the branch of the projectable Ho\v{r}ava--Lifshitz gravity with the ghosts in the IR is
a phenomenologically viable scenario. We note, however, that the presence of the
strong coupling by itself is not good from the viewpoint of renormalizing Ho\v{r}ava--Lifshitz gravity\footnote{In particular, the presence of the strong coupling invalidates
the discussion of Ref.~\cite{Barvinsky:2015kil}, where it is assumed that the model can be treated perturbatively at the arbitrary scale.}. Naively, one could circumvent the problem by setting the UV cutoff
of the theory somewhat below the would be strong coupling scale. In that case, the
computation of the strong coupling cutoff performed within the IR theory is
not valid. Instead, one can argue that the theory remains perturbative at an arbitrary scale~\cite{Blas:2009ck}. However, this is not a viable option in the branch with the positive sound speed squared. The reason is that the ghost
instabilities are not cured by the operators arising in the UV. Hence, retaining perturbativity would make them propagate at an arbitrary scale and lead to an instantaneous destabilization of the vacuum.
Thus, we are forced to assume that the UV cutoff is above the strong
coupling scale, and hence the projectable Ho\v{r}ava--Lifshitz gravity
enters the genuinely non-perturbative regime at some point. This is
a non-appealing feature of the IR theory, which severely obscures
its UV completion. We sketch a possible solution of the problem
in the end of Subsection 2.2 and in Appendix. This solution, however, involves operators beyond the projectable
Ho\v{r}ava--Lifshitz gravity and, thus, is out of the main scope of the paper.

Apart from the issues with the UV completion,
the scenario at hand looks attractive from a somewhat more down-to-earth perspective: it allows us to construct a fluid free of any obvious pathologies characterized by a small sound speed. In particular, that fluid could be an interesting candidate for Dark Matter. While practically indistinguishable from a collection of cold non-interacting particles at the background and linear levels~\cite{Mukohyama:2009mz, Chamseddine:2013kea, Chamseddine:2014vna, Capela:2014xta, Mirzagholi:2014ifa}, it
exhibits a different behaviour in the non-linear phase~\cite{Mukohyama:2009tp, Capela:2014xta}.
These features might be of sufficient interest in view of the series of
small scale problems alleged to the Cold Dark Matter~\cite{Weinberg:2013aya}.

Finally, we note that the results of the present paper can be literally translated into the context of the mimetic
matter scenario~\cite{Chamseddine:2013kea}. The latter is a novel proposal, originally designed to explain the Dark Matter in the Universe
by a singular disformal transformation of the metric in GR~\cite{Deruelle:2014zza}. In fact, the mimetic matter scenario extended by means
of a higher derivative term~\cite{Chamseddine:2014vna} is equivalent to the IR limit of the projectable Ho\v{r}ava--Lifshitz gravity, as well as to a particular
version of the Einstein-Aether theory~\cite{Jacobson:2000xp, Jacobson:2014mda, Haghani:2014ita}. The classical evolution of cosmological perturbations in that setup was studied in Refs.~\cite{Chamseddine:2014vna, Capela:2014xta}, and the stability issues have been partially addressed in Ref.~\cite{Barvinsky:2013mea}. In a certain sense, the results of the present paper extend those analyses to the quantum level.

The paper is organized as follows. In Section~2, we review the IR limit of the Ho\v{r}ava--Lifshitz gravity
with the projectability condition imposed and re-derive the strong coupling scale.
In Section~3, we estimate the decay rate of the vacuum into
photons and ghosts and establish an upper limit
on the sound speed $c_s$ and, consequently, on the
scale of the strong coupling. Finally in Section~4, we discuss some prospects for future research.

\section{Review of projectable Ho\v{r}ava--Lifshitz gravity}

\subsection{Setup}
We start with a brief review of the Ho\v{r}ava--Lifshitz gravity theory, focusing on its projectable version. We write the metric using the Arnowitt--Deser--Misner (ADM) formalism~\cite{ADM},
\begin{equation}
\nonumber
ds^2= N^2 dt^2-\gamma_{ij} (dx^{i}+N^{i}dt) (dx^{j} +N^{j}dt) \; ,
\end{equation}
(we assume the mostly negative signature). Along with Eq.~\eqref{scaling}, one postulates the following scaling of the lapse function $N$, the shift vector $N_i$ and the metric on the
constant $t$-hypersurface $\gamma_{ij}$~\cite{Horava:2009uw},
\begin{equation}
\nonumber
\gamma_{ij} \rightarrow \gamma_{ij}\;, \quad N_i \rightarrow b^2N_i\;, \quad N \rightarrow N \; .
\end{equation}
Here the value of the critical exponent $z=3$ is implied. Next, one classifies all the possible operators according to their scaling dimensions. In particular, the operators entering
the GR kinetic term, i.e.,
\begin{equation}
\nonumber
K_{ij}K^{ij} \; , \qquad K^2\;,
\end{equation}
have scaling dimension $6$. Here $K_{ij}$ and $K$ are the extrinsic curvature tensor and its trace,
respectively. The operators of the same dimension as the kinetic terms are called marginal. The action of the Ho\v{r}ava--Lifshitz gravity is then
designed as the sum of all marginal and relevant operators~\cite{Horava:2009uw}. These can be assembled in the GR fashion,
\begin{equation}
\label{Horava}
S=\frac{1}{16\pi G}\int dt d^3x \sqrt{\gamma} N\left(K_{ij}K^{ij}-\lambda K^2 - {\cal V}\right) \; ,
\end{equation}
where ${\cal V}$ is the so called potential typically involving powers of the 3-dimensional Ricci
scalar and tensor as well as their derivatives~\cite{Horava:2009uw, Sotiriou:2009gy, Sotiriou:2009bx}. Just to illustrate the main idea behind
Ho\v{r}ava's proposal, let us write down an example of an operator entering the potential ${\cal V}$,
\begin{equation}
\label{margrel}
 \frac{1}{M^4_*}R\Delta R \; .
\end{equation}
Here $R$ is the Ricci scalar calculated on the constant $t$-hypersurface; $M_*$ is a free parameter defining the scale of the UV completion of the theory.
In the presence of the marginal operator~\eqref{margrel}, the graviton propagator gets modified as follows~\cite{Horava:2009uw},
\begin{equation}
\label{prop}
\frac{1}{\omega^2 -{\bf k}^2} \rightarrow \frac{1}{\omega^2 -{\bf k}^2 -\frac{({\bf k}^2)^3}{M^4_*}} \; .
\end{equation}
That is, at sufficiently long distances, one recovers the standard GR behaviour of the
propagator. At high energies, the scaling of the propagator strongly improves. Consequently, one opens up the
possibility to reduce the divergencies of the loop integrals. In particular,
 the behaviour as in Eq.~\eqref{prop} leads to the power counting
renormalizable gravity in 4 dimensions. 

Having briefly described the UV properties of the Ho\v{r}ava--Lifshitz
gravity, we switch to its IR limit,---the main focus of our studies.
In that limit, the potential ${\cal V}$ takes the form\footnote{Generically, one should write ${\cal V} \rightarrow -\zeta R$, where $\zeta$ is some arbitrary constant. This, however, can be safely tuned to unity to better match the GR case.},
\begin{equation}
\label{potentialir}
{\cal V} \rightarrow -R \; .
\end{equation}
Substituting this into the action~\eqref{Horava}, we obtain the action of GR, modulo the constant $\lambda$, which is generically not equal to unity.
Naively, Einstein's gravity is recovered in the limit $\lambda \rightarrow 1$. However, this is not so. The reason is that the action of Ho\v{r}ava--Lifshitz gravity respects a smaller group
of diffeomorphisms compared to that of GR~\cite{Horava:2009uw},
\begin{equation}
\nonumber
t \rightarrow \tilde{t}(t)\;, \qquad {\bf x} \rightarrow \tilde{{\bf x}} (t, {\bf x}) \; ,
\end{equation}
called foliation-preserving diffeomorphisms. This eventually leads to an additional
degree of freedom on top of the helicity-2 graviton. The new degree of freedom
typically exhibits a pathological behaviour in the IR or leads to the breakdown of perturbation theory at unacceptably large distances~\cite{Charmousis:2009tc, Blas:2009yd}. So, unlike in GR, in Ho\v{r}ava--Lifshitz gravity
(or, at least, in its original incarnation) the lapse $N(t, {\bf x})$ is a dynamical variable. This variable
is plagued by gradient instabilities and a very low strong coupling scale.

That problem is absent in the projectable version of the Ho\v{r}ava--Lifshitz gravity~\cite{Horava:2009uw}, to which we turn now. There, one assumes that the lapse is a
function of time only, i.e., $N=N(t)$. In particular, it can be set to unity, i.e., $N(t)=1$,
by the proper choice of the time reparametrization. To enforce the condition $N=1$, following Ref.~\cite{Blas:2009qj}, we introduce a term in the action with a
Lagrange multiplier $\Sigma$,
\begin{equation}
\label{fix}
S_{fix}=\int d^3x dt \sqrt{\gamma}N \frac{\Sigma}{2} \left(\frac{1}{N^2}-1 \right) \; .
\end{equation}
This is not the only option. Instead, one may consider a ghost condensate-like term~\cite{Blas:2010hb},
i.e., $M^4 (1/N^2-1)^2$, where $M$ is a fictitious mass parameter. The projectability condition is then ensured by imposing the limit $M \rightarrow \infty$.
In what follows, we will see that the two strategies lead to the same conclusions. Combining Eqs.~\eqref{Horava},~\eqref{potentialir} and~\eqref{fix}, we are ready to write down the
action for the IR limit of the projectable Ho\v{r}ava--Lifshitz gravity,
\begin{equation}
\label{HLinf}
S_{IR}=S_{GR}+\int d^3x dt \sqrt{\gamma}N \left[\frac{\Sigma}{2} \left(\frac{1}{N^2}-1 \right)-\frac{\lambda-1}{16\pi G} K^2\right] \; .
\end{equation}
We explicitly grouped a part of the terms into the GR action $S_{GR}$.

Despite the absence of the problems with the lapse, the model~\eqref{HLinf} has a propagating helicity-0 mode,
which exhibits ghost or gradient instabilities depending on the value of the parameter $\lambda$~\cite{Blas:2010hb, Koyama:2009hc}. To study the properties of that mode is the main goal of the present paper.
For this purpose, the action~\eqref{HLinf} is not very convenient. Instead, we choose to work with
an equivalent formulation of Eq.~\eqref{HLinf}, which is manifestly invariant under the full GR diffeomorphisms. This is achieved by performing the Stuckelberg
trick, with the Stuckelberg field $\varphi$ dubbed as {\it khronon} in the context of the Ho\v{r}ava--Lifshitz gravity~\cite{Blas:2009yd, Blas:2009qj, Blas:2010hb, Germani:2009yt, Kimpton:2010xi}. Note that by introducing the field
$\varphi$, one does not enlarge the number of degrees of freedom in the theory, but singles
out those already present. In particular, the dynamics of the field $\varphi$ is eliminated
by imposing the unitary gauge, where it takes the form,
\begin{equation}
\nonumber
\varphi=t \; .
\end{equation}
We see that the field $\varphi$ defines the absolute time. Hence, the
name 'khronon'.

To write the action~\eqref{HLinf} in covariant form, it is enough to promote
the lapse $N$ and the extrinsic curvature $K_{\mu \nu}$ to the quantities invariant under GR
diffeomorphisms. This has been done in Ref.~\cite{Blas:2009yd}, and we omit the details of the calculations here. The result for the lapse $N$ reads,
\begin{equation}
\label{lapsestuck}
\frac{1}{N^2} \rightarrow  g^{\mu \nu} \partial_{\mu} \varphi \partial_{\nu} \varphi \; .
\end{equation}
Now $K_{\mu \nu}$ is the extrinsic curvature of the
hypersurface $\varphi=$const, and its trace is given by
\begin{equation}
\label{K}
K \rightarrow \nabla_{\mu} \left(\frac{\nabla^{\mu} \varphi}{\sqrt{g^{\mu \nu} \partial_{\mu }\varphi \partial_{\nu} \varphi}} \right) \; .
\end{equation}
This expression can be simplified by virtue of the constraint enforced
by the Lagrange multiplier $\Sigma$. In the Stuckelberg
treatment the constraint is given by,
\begin{equation}
\label{constraint}
g^{\mu \nu} \partial_{\mu} \varphi \partial_{\nu} \varphi=1 \; .
\end{equation}
This can be safely plugged into Eq.~\eqref{K} readily at the level of the
action. See Appendix~A for the justification. As a result, Eq.~\eqref{K} gets simplified, 
\begin{equation}
\nonumber
K \rightarrow \square \varphi \; .
\end{equation}
The covariant action for the IR limit of the projectable Ho\v{r}ava--Lifshitz gravity is given by~\cite{Blas:2010hb},
\begin{equation}
\label{IDM}
S_{IR}=S_{GR}+\int d^4 x \sqrt{-g} \left[\frac{\Sigma}{2}\left(g^{\mu \nu} \partial_{\mu} \varphi \partial_{\nu} \varphi -1 \right)+\frac{\gamma}{2} (\square \varphi )^2 \right] \; ,
\end{equation}
where we introduced the shorthand notation,
\begin{equation}
\frac{1-\lambda}{8\pi G} = \gamma \; .
\end{equation}
In passing, it is interesting to note that the action~\eqref{IDM} by itself is not 
specific to the projectable Ho\v{r}ava--Lifshitz gravity, but may arise
in a drastically different framework. In this regard, the mimetic
matter scenario has brought some attention
recently~\cite{Chamseddine:2013kea}. The idea is to consider a
particular (singular) conformal transformation of the metric, i.e., $g_{\mu \nu} \rightarrow
\tilde{g}_{\mu \nu}= g^{\alpha \beta} \partial_{\alpha}
\varphi \partial_{\beta} \varphi \cdot g_{\mu \nu}$. Here $\varphi$ is
some scalar field.
This transformation  does not leave the GR equations of motion invariant.
The discrepancy from GR is equivalent to extending the Einstein-Hilbert action by means of the term with the Lagrange
multiplier as in Eq.~\eqref{IDM}~\cite{Barvinsky:2013mea,Golovnev:2013jxa,Hammer:2015pcx}. There is, however, a conceptual distinction
from the Ho\v{r}ava--Lifshitz model.
In the mimetic matter case, the higher derivative
term as in Eq.~\eqref{IDM} is added in view of some phenomenological goals~\cite{Chamseddine:2014vna, Capela:2014xta, Mirzagholi:2014ifa}, i.e., it does not follow immediately
from the first principles underlying the scenario.

Keeping in mind this potentially interesting scenario, we proceed with the Ho\v{r}ava--Lifshitz model as the main focus
of the present work. The action~\eqref{IDM} will be the starting point of our further discussions.

\subsection{Low-energy quadratic action}

Interestingly, at the level of the background cosmological equations, the
model given by the action~\eqref{IDM} describes dust (pressureless perfect fluid).
This readily follows from the $ij$-components of Einstein's equations~\cite{Capela:2014xta,
Mirzagholi:2014ifa, Haghani:2014ita},
\begin{equation}
\label{ij}
{2\cal H}'+{\cal H}^2=0 \; .
\end{equation}
The $00$-component of Einstein's equations is given by
\begin{equation}
\label{00}
3{\cal H}^2 (1-24\pi \gamma G)= 8\pi Ga^2 \bar{\Sigma} \; .
\end{equation}
Here a prime denotes the derivative with respect to the
conformal time, ${\cal H} \equiv  a'/a$, $a$ is the scale factor, and ${\bar \Sigma}$ is the
background value of the Lagrange multiplier field. As it follows from Eqs.~\eqref{ij} and~\eqref{00}, at the background level the higher derivative term is irrelevant: effects due to the
non-zero parameter $\gamma$ can be absorbed into the redefinition of the
cosmological Newton's constant. That degeneracy with the case of the pressureless fluid gets broken at the linear level,
where the $\gamma$-term gives rise to a non-zero scalar sound speed as
will be clear  shortly.
Before that, let us point out an intriguing possibility to mimic the energy density of the
Dark Matter by the field $\Sigma$, i.e., without invoking physics beyond the Standard Model of particles. In this picture, the khronon $\varphi$ is understood as the velocity potential, while the constraint~\eqref{constraint} leads to the geodesics equation followed by the dust particles in the gravitational field. The
properties of the Dark Matter in the model~\eqref{IDM} and closely related scenarios have been explored in Refs.~\cite{Mukohyama:2009mz, Mukohyama:2009tp} in the context
of the projectable Ho\v{r}ava--Lifshitz gravity, in Refs.~\cite{Chamseddine:2013kea, Chamseddine:2014vna, Capela:2014xta, Mirzagholi:2014ifa, Barvinsky:2013mea, Golovnev:2013jxa, Hammer:2015pcx,  Arroja:2015wpa, Arroja:2015yvd} in the mimetic matter setup and in Ref.~\cite{Lim:2010yk} on purely phenomenological grounds\footnote{Note that Dark Matter is the prediction specific to the projectable version of the Ho\v{r}ava--Lifshitz model. Still, in the different extensions of the model, one can entertain the opportunity of having MOND-like phenomenology in the IR limit~\cite{Blanchet:2011wv, Bonetti:2015oda}.}.  That line
of research is out of the scope of the present paper.

Naively, the higher derivative term in the action~\eqref{IDM} would lead to an
Ostrogradski instability. However, the Ostrogradski theorem
is not applicable to models with constraints~\cite{Chen:2012au, Woodard1, Woodard2} (the so-called degenerate systems). The reason is that the addition of the constraints may reduce the dimension of the Hamiltonian phase space, and consequently, the number of the propagating
degrees of freedom may be smaller compared to the naive counting. This is indeed the case in the
projectable Ho\v{r}ava--Lifshitz gravity. In particular, the scalar sector of the model has only one propagating degree of freedom.

To study linear perturbations in the projectable Ho\v{r}ava--Lifshitz model, we use the standard conventions~\cite{Mukhanov:1990me, Gorbunov:2011zzc},
\begin{equation}
\nonumber
N=a\left(1+\Phi \right)\;, \quad N_i =a^2 \partial_i B\;, \quad \gamma_{ij} =a^2 (1-2\Psi) \delta_{ij}
+2a^2 \partial_{i}\partial_jE \; .
\end{equation}
Here $\Phi, \Psi, B, E$ are the scalar potentials, and
we omit the discussion of the vector and tensor modes (see the comment at the end
of this Subsection).

At the linear level, the constraint $g^{\mu \nu} \partial_{\mu} \varphi \partial_{\nu} \varphi=1$  gives
\begin{equation}
\label{linc}
\frac{\delta \varphi'}{a}=\Phi \; .
\end{equation}
The latter is enforced by the Lagrange multiplier field $\Sigma$.  
Therefore, one can safely use  Eq.~\eqref{linc} to eliminate the field $\Phi$ without affecting the dynamics of the remaining degrees of freedom~\cite{Pons}. Upon  substituting Eq.~\eqref{linc}  in the quadratic action, integrating by parts, dropping boundary terms and making use of the background equations~\eqref{ij} and~\eqref{00}, we obtain
\begin{equation}
\label{action}
\begin{split}
\delta_2 S_{IR} &= \frac{1}{16\pi G} \int d^4 x a^2 \Bigl[-6{\cal R}'^2+2\Delta \Psi \Bigl(2\frac{\delta \varphi'}{a}-\Psi\Bigr)
+\frac{3{\cal H}^2\delta \varphi \Delta \delta \varphi}{a^2}  +4{\cal R}' \cdot (\Delta E'-\Delta B)\Bigr]+\\
&\quad+\frac{\gamma}{2}\int d^4x a^2  \Bigl(3{\cal R}'+ \frac{\Delta \delta \varphi}{a}+ \Delta B-\Delta E' \Bigr)^2 \; .
\end{split}
\end{equation}
Here ${\cal R}$ is the gauge-invariant curvature perturbation, defined as ${\cal R}=\Psi +\frac{\cal H}{a}\delta \varphi$.
Note that the action~\eqref{action} does not assume any gauge choice. The fields $B$ and $E$ enter only via the combination $B-E'$, as it
should be, because of diffeomorphism invariance. The variation with respect to the field $B$ (or $E'$) yields
\begin{equation}
\label{constr}
\Delta B -\Delta E'= \frac{1}{4\pi \gamma G} {\cal R}' -3{\cal R}'-\frac{\Delta \delta \varphi}{a} \; .
\end{equation}
At this level, we explicitly assume that the parameter $\gamma$ is non-zero, i.e., $\gamma \neq 0$. As it follows from Eq.~\eqref{constr},
$B$ is an auxiliary field. Namely, it is separated from the other fields by means
of its own equation of motion. Therefore, one can safely substitute the constraint~\eqref{constr} into
the quadratic action~\eqref{action}~\cite{Pons}. Doing so, we obtain
\begin{equation}
\label{eff}
\delta_2 S_{IR} = \frac{1}{8\pi G} \int d^4 x a^2\left(-\frac{1}{4\pi \gamma G}{\cal R}'^2 +3{\cal R}'^2-
{\cal R} \Delta {\cal R}  \right) \; .
\end{equation}
As it was expected, there is only one propagating degree of freedom in the scalar sector of the model.
Furthermore, in the unitary gauge $\delta \varphi =0$, one reproduces the result of Ref.~\cite{Blas:2010hb}. This serves as a
simple cross-check of our calculations.

Note that the action~\eqref{eff} has a continuous limit to the flat space-time. This allows us to set consistently $a \rightarrow 1$ and ${\cal H} \rightarrow 0$, and ignore the effects related to the expansion of the Universe in what follows. These have been already discussed to some extent in Refs.~\cite{Mukohyama:2009mz, Chamseddine:2013kea, Chamseddine:2014vna, Capela:2014xta, Mirzagholi:2014ifa, Lim:2010yk}. In the Minkowski limit, the gauge-invariant variable ${\cal R}$ takes the form ${\cal R}=\Psi$, and
the action~\eqref{eff} reduces to,
\begin{equation}
\label{psi}
\delta_2 S_{IR}= \frac{1}{8\pi G}\int d^4x \left( -\frac{1}{c^2_s} \dot{\Psi}^2 -\Psi \Delta \Psi \right) \; .
\end{equation}
Here $c^2_s$ is the sound speed squared given by~\cite{Blas:2009qj, Blas:2010hb, Koyama:2009hc},
\begin{equation}
\nonumber
c^2_s =\frac{4\pi \gamma G}{1-12\pi \gamma G} \; .
\end{equation}
As it follows, the IR properties of the field $\Psi$ are characterized by the phonon-like dispersion relation
\begin{equation}
\label{disp}
\omega^2 =c^2_s {\bf p}^2 \; .
\end{equation}
Hereafter, we assume the hierarchy $\gamma \ll M^2_{Pl}$, which allows us to
simplify the expression for the sound speed squared $c^2_s$,
\begin{equation}
\label{sound}
c^2_s = 4\pi \gamma G \; .
\end{equation}
This is, in fact, the
only phenomenologically viable option, as it
will be clear from the following discussions.

Let us start with the case $\gamma = 0$. This corresponds to GR supplemented by a pressureless perfect fluid (dust), which is a classically well-defined system (up to the caustic singularities).  On the other hand,
the action~\eqref{psi} is ill-defined in that case. However, this does not
signal the inconsistency in the discussion, as we assumed the choice $\gamma \neq 0$ at the intermediate step (see the comment
after Eq.~\eqref{constr})\footnote{Of course, one can rewrite Eq.~\eqref{constr} in the form applicable for both cases $\gamma =0$ and $\gamma \neq 0$, i.e.,
$\gamma (\Delta B -\Delta E')= \frac{1}{4\pi G} {\cal R}' -3\gamma {\cal R}'-\gamma \frac{\Delta \delta \varphi}{a} $. Still, the choice $\gamma=0$ is 'singular' in a sense that it does not
allow to integrate out the field $B$, which plays the role of the Lagrange multiplier now. }. One obvious way to handle the situation is to consider the limit of the infinitely small parameter $\gamma$, i.e.,
$\gamma \rightarrow 0$, instead of setting it exactly to zero. Then, demanding that the action~\eqref{psi} (\eqref{eff}) remains finite, one obtains the first order equation for the potential $\Psi$ (the curvature perturbation ${\cal R}$): $\dot{\Psi}=0$ (${\cal R}'=0$). The latter is recognized
as the conservation of the potential (curvature perturbation) characteristic of
dust. There is a more trustworthy way to get the same equation. That is,
one gets back to the original quadratic action~\eqref{action}, which is manifestly applicable for the
arbitrary values of the parameter $\gamma$. Then, the conservation of the curvature perturbation follows immediately upon varying with respect to the
field $B$. Furthermore, one may check that all the other equations following from Eq.~\eqref{action} (with $\gamma=0$ understood) are the same as in GR supplemented by a pressureless perfect fluid. Although the simple dust model is typically employed to describe the behaviour of Dark Matter on cosmological scales, it has two important drawbacks rendering the case $\gamma \rightarrow 0$ pathological. First,
the quantum properties of the model are unclear, as the strong coupling scale 
tends to zero in that limit~\cite{Blas:2010hb, Koyama:2009hc}\footnote{Alternatively, the problem with the quantization can be understood
from the fact that the curvature perturbation satisfies the first order equation of motion in the limit $c_s \rightarrow 0$.}. See the discussion in the next Subsection. Second, a pressureless perfect fluid develops caustic singularities at a finite time~\cite{Landau}.

Therefore, we switch to the case $\gamma \neq 0$ in what follows. For negative values of the parameter $\gamma$ (sound speed squared), the field $\Psi$ suffers from gradient instabilities. On the other hand, positive values of the parameter $\gamma$ (sound speed squared), lead to the 'wrong' sign of the kinetic term in the action~\eqref{psi}. The study of this ghost-unstable branch of the projectable Ho\v{r}ava--Lifshitz gravity will be our primary interest in the present paper.

Note one important difference between
ghost and gradient instabilities in the projectable
Ho\v{r}ava--Lifshitz gravity. At sufficiently high spatial
momenta, the action~\eqref{psi} must be completed by the relevant and marginal
operators encoded in the potential ${\cal V}$. In the unitary gauge, these result into quadratic terms of the form,
\begin{equation}
\label{LVHL}
\frac{1}{M^2_*}\Psi \Delta^2 \Psi\;, \qquad  \frac{1}{M^4_*}\Psi \Delta^3 \Psi \; .
\end{equation}
Recall that $M_*$ is the scale at which the Lorentz-violating operators
become important presumably renormalizing gravity
in the UV. As it follows from the
structure of the terms~\eqref{LVHL}, they are capable to cure gradient instabilities in the UV, while leaving the sign
of the kinetic term in Eq.~\eqref{psi} intact. Hence, from the higher derivative perspective alone, ghosts are unavoidable
in the scenario with the positive parameter $\gamma$. This situation may change due to the presence of sufficiently
low strong coupling scale, above which the theory is hopefully
free of ghost/gradient instabilities.

This problem with the ghosts, we note, is specific to the projectable Ho\v{r}ava--Lifshitz model,
and can be avoided by relaxing some assumptions underlying the framework. Here is a sketch of
one possible solution. Let us assume an extension of the Ho\v{r}ava--Lifshitz
gravity by means of the operator,
\begin{equation}
\label{higherdim}
\frac{1}{M^2_*} K \Delta K \; .
\end{equation}
Recall that $K$ is the trace of the extrinsic curvature tensor. Generically, the operators of the form~\eqref{higherdim}  may compromise power counting renormalizability of the Ho\v rava--Lifshitz gravity~\cite{Colombo, Colombo1, Coates}. Therefore, 
they have been omitted in the original action~\eqref{Horava}. On the other hand,  the term~\eqref{higherdim} and similar ones are quite well motivated, as 
they allow to stabilize the percolation of the Lorentz-violating effects 
from the gravity sector to the particle one~\cite{Pospelov}. 

In the Stuckelberg treatment, the operator~\eqref{higherdim} can be rewritten as follows
%
\begin{equation}
 \label{Stuck}
 \frac{1}{M^2_*}\square \varphi (\square -\partial_{\mu} \varphi \partial_{\nu} \varphi \nabla^{\mu}
 \nabla^{\nu})  \square \varphi \; .
 \end{equation}
Introducing this term into the action~\eqref{IDM} indeed allows us to recover the
positive sign of the kinetic term of the potential $\Psi$ in the UV, i.e., for spatial
momenta $|{\bf p}| \gtrsim M_*$. We relegate the details of the computations to the
Appendix B, and postpone a more thorough analysis for future work.

One comment is in order here. Compared to the
scalar sector, the tensor part of the projectable Ho\v{r}ava--Lifshitz gravity exhibits healthy behaviour in the IR limit.
Namely, it is free of ghost/gradient instabilities and the strong coupling issues. See, e.g., Eq.~(9) of Ref.~\cite{Barvinsky:2015kil}.  Nevertheless, the interaction with the strongly coupled
scalar sector may essentially affect the behaviour of the tensor modes at larger momenta,
severely obscuring their UV properties.

\subsection{Cubic interactions: determining the strong coupling scale}

The strong coupling scale has been calculated previously in Refs.~\cite{Blas:2010hb, Koyama:2009hc} and is given by Eq.~\eqref{str}. In the present Subsection, we re-derive this result by employing the Newtonian gauge.

To understand the structure of the cubic interactions, we expand the constraint equation~\eqref{linc} up to the
quadratic terms in the potential $\Phi$ and the khronon field perturbations $\delta \varphi$\footnote{Recall that we choose to work in the Minkowski background.},
\begin{equation}
\label{constrquadr}
\Phi=\delta \dot{\varphi}+\frac{1}{2}\delta \dot{\varphi} \delta \dot{\varphi}- \frac{1}{2}\partial_{i} \delta\varphi \partial_i \delta \varphi \; .
\end{equation}
Note that the terms $\sim \Phi^2 $ and $\sim \delta \dot{\varphi}\cdot  \Phi$ cancel out upon implementing
the first order constraint $\delta \dot{\varphi}=\Phi$. In the Newtonian gauge, the strong coupling stems from the
following term in the GR action,
\begin{equation}
\label{danger}
\sim M^2_{Pl} \Delta \Psi \Phi\;,
\end{equation}
(cf. the second term in the first line of Eq.~\eqref{action}). This is quadratic in the fields $\Psi$ and $\Phi$, and, thus,
naively does not correspond to any interaction. According to Eq.~\eqref{constrquadr}, however,
the potential $\Phi$ is sourced by the quadratic order perturbations in the khronon field. Therefore, we are left with the
following interaction,
\begin{equation}
\label{cube}
\sim M^2_{Pl} \Delta \Psi \partial_i \delta \varphi \partial_i \delta \varphi \; .
\end{equation}
We ignored the terms with the time derivatives in Eq.~\eqref{constrquadr}.
This is legitimate in view of the dispersion relation $\omega^2 = c^2_s {\bf p}^2$, where
$c^2_s \ll 1$.
The field $\delta \varphi$ is extracted from the constraint~\eqref{constr}, where we set the
fields $B$ and $E$ to zero by the Newtonian gauge choice,
\begin{equation}
\nonumber
\delta \varphi \sim \frac{M^2_{Pl}\dot{\Psi}}{\gamma \Delta } \; .
\end{equation}
Substituting this into Eq.~\eqref{cube}, we get an estimate for the cubic interaction,
\begin{equation}
\nonumber
\sim \frac{M^6_{Pl} \Psi}{\gamma^2 \Delta} ( \partial_i \dot{\Psi})^2 \; .
\end{equation}
Again, taking into account the dispersion relation~\eqref{disp} and Eq.~\eqref{sound},
one rewrites the estimate above as follows,
\begin{equation}
\nonumber
\sim \frac{M^4_{Pl}}{\gamma }\Psi (\partial_i \Psi)^2 \; .
\end{equation}
This is to be compared with the standard quadratic term in the GR action involving the spatial derivatives of the potential $\Psi$\footnote{More rigorously,
to deduce the strong coupling scale, one calculates the cross-section of the scattering of two $\Psi$-particles. The obtained cross-section should not violate the optical
theorem, i.e., unitary must be obeyed. This gives a constraint on the allowed values of the
momenta. We followed this way, and showed that the result matches the one given below.},
\begin{equation}
\label{est}
\frac{\frac{M^4_{Pl}}{\gamma }\Psi (\partial_i \Psi)^2}{M^2_{Pl} \Psi \Delta \Psi} \sim \frac{M^2_{Pl}}{\gamma} \Psi  \; .
\end{equation}
To proceed, we need an estimate for the amplitude of the fluctuations of the field $\Psi$.
For this purpose, we switch to the canonical normalized variable $\hat{\Psi}= \frac{M^2_{Pl}}{\sqrt{\gamma}} \Psi$.
Fluctuations of the variable $\hat{\Psi}$ are characterized by a
Gaussian distribution with zero mean value and the variance,
\begin{equation}
\nonumber
\langle \hat{\Psi}^2 \rangle \sim \int \frac{d{\bf p}}{|\omega ({\bf p})|} \sim \frac{M_{Pl}}{|\gamma|^{1/2}} |{\bf p}|^2\; .
\end{equation}
Returning to the variable $\Psi$, we obtain the estimate for its fluctuations,
\begin{equation}
\nonumber
\Psi \sim \sqrt{\langle \Psi^2 \rangle} \sim \frac{|\gamma|^{1/4} |{\bf p}|}{M^{3/2}_{Pl}} \; .
\end{equation}
Combining everything together and demanding that the ratio~\eqref{est} does not exceed unity,
we conclude that the scale of the strong coupling is
\begin{equation}
\label{cutoffp}
\Lambda_p \sim \frac{|\gamma|^{3/4}}{M^{1/2}_{Pl}} \; .
\end{equation}
This result exactly matches the one obtained in Refs.~\cite{Blas:2010hb, Koyama:2009hc}.
In particular, the scale of strong coupling tends to zero in two limits:
in the decoupling limit ($M_{Pl} \rightarrow \infty$) and
in the limit of the pressureless perfect fluid ($\gamma \rightarrow 0$).
Note that the cutoff~\eqref{cutoffp} is applied to the spatial momenta only, i.e.,
it breaks Lorentz-invariance explicitly. Hence, the subscript '$p$'. To understand the region of energies
which can be treated perturbatively, one simply makes use of the dispersion relation~\eqref{disp}.
This yields,
\begin{equation}
\label{cutoffw}
\Lambda_{\omega} \sim \frac{|\gamma |^{5/4}}{M^{3/2}_{Pl}} \; .
\end{equation}
That cutoff is not of particular importance in the projectable
Ho\v{r}ava--Lifshitz gravity, but merely reflects the fact that the quanta of the
field $\Psi$ are 'slow'. Indeed, the UV completing operators written schematically in Eq.~\eqref{LVHL} carry only spatial derivatives, and thus, are sensitive
only to the scale~\eqref{cutoffp}.

For the model to be phenomenologically viable, the strong coupling
scale $\Lambda_p$ must be larger than the maximal scale, at which GR has been tested, i.e.,
\begin{equation}
\label{allowed}
\Lambda_p\gtrsim 10^{-3}~\mbox{eV} \; .
\end{equation}
This translates into the bound on the parameter $\sqrt{|\gamma|}$,
\begin{equation}
\label{gammalower}
\sqrt{|\gamma|} \gtrsim 10~\mbox{MeV} \; .
\end{equation}
In the branch of the projectable Ho\v{r}ava--Lifshitz gravity plagued by the
gradient instabilities at low spatial momenta, the
associated constraints are orders of magnitude stronger~\cite{Blas:2010hb},
making it phenomenologically non-viable. Let us show this explicitly. Due to the presence of the Lorentz-violating terms
as in Eq.~\eqref{LVHL}, gradient instabilities are cut at the scale $|{\bf p}| \sim M_*$.
One then  demands that the time at which they propagate exceeds the age of the Universe.
Namely, $|\mbox{Im} ~\omega |\lesssim |c_s| M_* \lesssim H_0$, where
$H_0$ is the Hubble constant. The parameter $M_*$ is bounded from below, $M_* \gtrsim 10^{-3}$ eV,---otherwise, Lorentz violating effects would pop out at sub-mm scales, in conflict with GR tests.
Combining everything together, we obtain the constraint on the strong coupling scale: $\Lambda_p \lesssim 10^{-17}$ eV~\cite{Blas:2010hb}.
This is by many orders of magnitude lower than the allowed value~\eqref{allowed}. Strictly speaking, having the hierarchy $\Lambda_p \ll M_*$, one cannot trust these results,
as they were obtained by exploiting the region of the momenta, where perturbation theory
breaks down. Instead, let us assume that gradient instabilities are cut
by the scale of the strong coupling itself---with the hope that the
theory is free of any instabilities in the non-linear regime. In that case, the range of
the momenta, where the theory can be treated perturbatively, is slightly extended: $\Lambda_p \lesssim 10^{-8}$ eV,---
still in conflict with the GR tests.

From this point on, we abandon the branch of the
Ho\v{r}ava--Lifshitz gravity characterized by the negative sound speed squared, and switch to the one plagued by the ghosts in the IR. In the end of the previous Subsection, we observed that the ghosts cannot be cured in the UV, at least in the
projectable Ho\v{r}ava--Lifshitz gravity, as it stands. In this regard, the presence of the
strong coupling is necessary to render the model phenomenologically acceptable.
Namely, above the scale $\Lambda_p$, the results of the linear theory are not valid anymore. Therefore, we may assume that the model is free of instabilities in
the non-perturbative regime. While looking quite speculative, this expectation has
some reasons behind it. Indeed, the term in the GR action, which is responsible
for the strong coupling, in the linear theory gives rise to the
kinetic term with the negative sign (see the second term in Eq.~\eqref{action}).

Accordingly to the discussion above, we must set the UV cutoff $M_*$ somewhat higher than
the scale $\Lambda_p$,
\begin{equation}
\label{cond}
M_* \gtrsim \Lambda_p \; .
\end{equation}
Otherwise, our conclusions about the scale of the strong coupling would not be legitimate.
Indeed, the estimate~\eqref{cutoffp} has been obtained within the
IR theory, and must be revisited, if the UV operators~\eqref{LVHL} become relevant
already in the weakly coupled regime.  In that case, one can argue that the model retains perturbativity at the arbitrary momenta~\cite{Blas:2009ck}. This inevitably causes the presence of the all-scale ghost instabilities, and, consequently, a catastrophically fast vacuum decay.

To summarize, by imposing the condition~\eqref{cond}, we sacrifice the
renormalization of the Ho\v{r}ava--Lifshitz gravity in favour of its phenomenological viability. Let us be not too pessimistic,
however. Indeed, apart from the projectable Ho\v{r}ava--Lifshitz model, the two issues,---low scale ghosts and strong coupling at large momenta,---are not necessarily
related to each other. In particular, introducing the higher dimension operators, as, e.~g., in Eq.~\eqref{higherdim}, one can simultaneously recover the positive sign
of the kinetic term of the potential $\Psi$, and retain perturbativity of the theory.
In the current work our goal is modest---to show that
the projectable Ho\v{r}ava--Lifshitz gravity in its original incarnation
is experimentally acceptable in the branch containing ghosts. Therefore, we postpone any detailed
investigation of this potentially interesting loophole
for the future.

\section{Vacuum decay}

Now, let us consider vacuum decay into ghosts and Standard Model particles. In the Lorentz-invariant theories of
gravity, the associated decay rate is infinite\footnote{This assumes that gravity
remains unmodified at all the scales. Given that standard gravity must be embedded
into some microscopic theory at the scales $\gtrsim M_{Pl}$, a more conservative
limit on the decay rate is $\Gamma \lesssim M^4_{Pl}$. This is nevertheless many
orders of magnitude larger than the phenomenologically allowed value.}. The
situation is different in the Lorentz-violating theories, given that there is a low energy cutoff $\Lambda$ on the $\it{spatial}$ momenta~\cite{Cline:2003gs}\footnote{On the other hand,
in the Lorentz-invariant theories, the cutoff is imposed on the center-of-mass energy $\sqrt{s}$ of the
colliding particles. In this situation, the region of the integration over the spatial momenta
is infinite. Namely, one can always boost the momenta, while keeping the quantity $\sqrt{s}$ unchanged.}.
If the ghosts interact with the matter only via gravity, the decay rate per space-time volume is typically estimated to be of the order
$\Gamma \sim \frac{\Lambda^8}{M^4_{Pl}}$. This is to be contrasted to the measured flux of $\mbox{MeV}$-photons~\cite{Sreekumar:1997un}, what yields the constraint on the cutoff scale
$\Lambda \lesssim 3$ MeV~\cite{Cline:2003gs}. The upper limit here is applied, provided
that the sound speed is of the order unity, i.e., there is no hierarchy between the
frequency and spatial momenta cutoffs. If $c_s^2\ll1$, then the constraint on the spatial momenta cutoff can be relaxed by several orders of magnitude.
The reason is twofold. First, there are kinematic considerations, which severely constrain
the phase space of the decay products for $c^2_s \ll 1$. Second, the Standard Model particles are coupled to the gravitational potential $\Psi$ (directly or via the graviton), which is different
from the canonically normalized variable $\hat{\Psi}$ by a huge factor, once again depending on the
sound speed. As a result, the
effective coupling between the matter particles and the field $\hat{\Psi}$ is very small. With all the factors taken together, the decay rate turns out to be suppressed by a large power
of the sound speed (compared to the case with the unit sound speed). In turn, this allows to extend the perturbative regime in the model to the TeV scales.

The discussion above is applied to the case where there is no direct interaction of the
khronon field with the Standard Model particles. We specialize to this case in what follows. That is, we assume that the khronon affects the matter properties only via the mixing
with the scalar gravitational potential $\Psi$. Apart from the vacuum stability
issues, by the direct coupling of the khronon to the matter, one risks to reprocess the Lorentz-violating effects to the
particle sector~\cite{Audren:2014hza, Blas:2012vn}\footnote{In fact, Lorentz violation percolates the particle sector even in 
the absence of the direct coupling of the khronon to the standard matter, i.e., via quantum gravity loops~\cite{Pospelov}.  However, this mechanism leads to the 
effects suppressed by the ratio of the UV cutoff $M_*$ and the Plank mass $M_{Pl}$.}. Typically, this is expected to modify the
dispersion relation of the particles leading to a potential conflict
with observational data~\cite{Jacobson:2005bg, Kostelecky:2008ts, Blas:2014aca, Liberati}.

\begin{figure}[h]
\centering
\scalebox{.8}{
\includegraphics{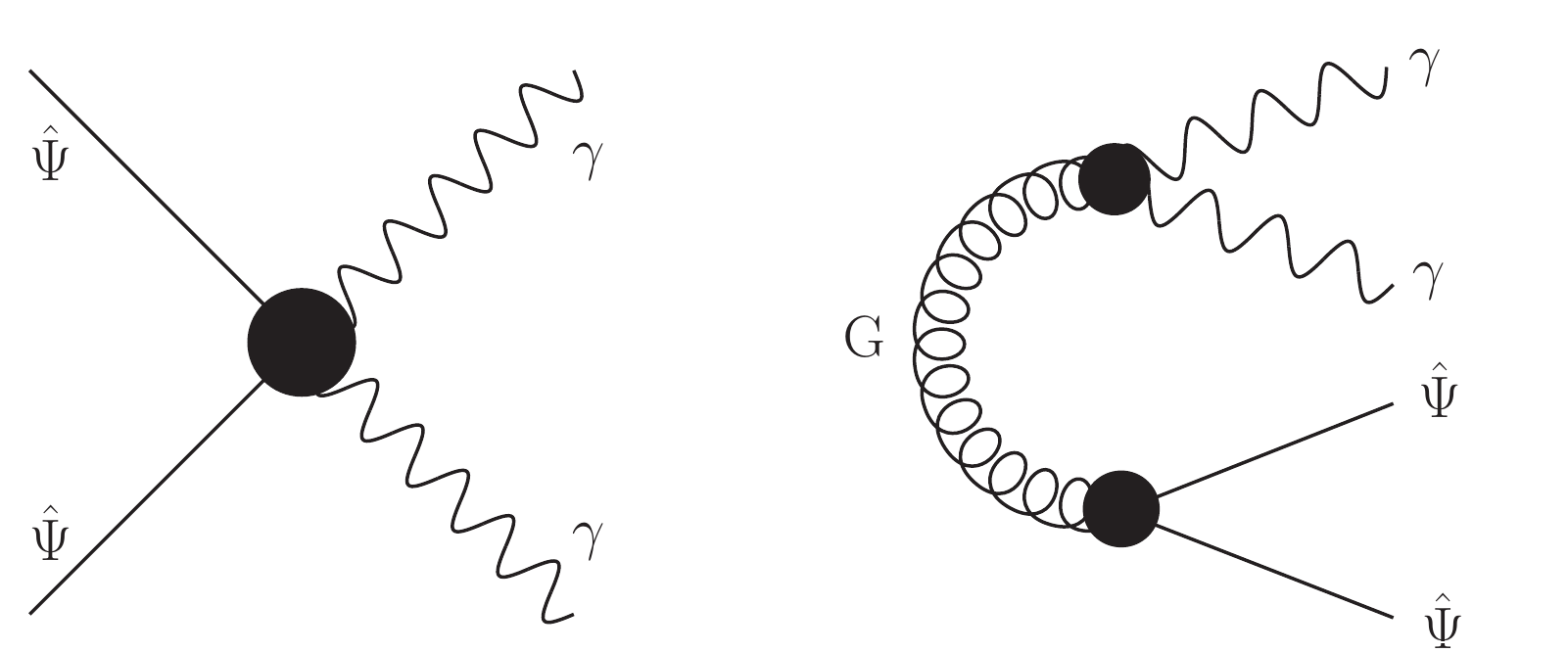}
}
\caption{Two diagrams of the vacuum decay into a pair of photons (the wiggly lines labelled by '$\gamma$') and a pair of ghosts (the straight lines labelled by '$\hat\Psi$'). The diagram on the left
follows from the 4-point contact interaction~\eqref{dir}. It gives the negligible
contribution to the total vacuum decay rate. The leading contribution comes from the diagram on the right involving the propagator of the graviton (the springy line labelled by 'G').}
\label{Fig:fig1}
\end{figure}

We will be primarily interested in the processes with photons in the
final state. These are argued to be the most relevant ones in Ref.~\cite{Cline:2003gs}. The action for electromagnetism is given by
\begin{equation}
\nonumber
S_{el}=-\frac{1}{4} \int d^4 x \sqrt{-g} \cdot  g^{\lambda \mu} g^{\rho \nu} F_{\mu \nu} F_{\lambda \rho} \; ,
\end{equation}
where $F_{\mu \nu} \equiv \nabla_{\nu} A_{\mu}-\nabla_{\mu} A_{\nu}=\partial_{\nu} A_{\mu}-\partial_{\mu} A_{\nu}$ is the electromagnetic tensor.
The interactions of the photons with the ghosts are of two sorts: those following from the
direct coupling of the photons to the scalar potential $\Psi$, and those involving
an exchange of a graviton. In terms of the canonically normalized field $\hat{\Psi}$, the former are described by\footnote{One
can write the analogous interactions but with the potential $\Phi$ instead of $\Psi$.
These, however, give nothing new, because of the relation $\Psi=\Phi$, which holds in linear theory.},
\begin{equation}
\label{dir}
{\cal L}_{ph-gh} \sim \quad \frac{\sqrt{\gamma}}{M^2_{Pl}} \cdot \hat{\Psi} \cdot F^2_{\mu \nu} \; , \quad  \frac{\gamma}{M^4_{Pl}} \cdot \hat{\Psi}^2 \cdot F^2_{\mu \nu} \; .
\end{equation}
Recall that the field $\hat{\Psi}$ is related to the scalar potential $\Psi$
by
\begin{equation}
\label{candef}
\hat{\Psi} \sim \frac{M^2_{Pl}}{\sqrt{\gamma}} \Psi \; .
\end{equation}
At the tree level, the first interaction term on the r.h.s. of~\eqref{dir} leads to the
process with two photons and one ghost particle in the final
state. Naively, this should be the dominant one. In fact, it does not occur
for the simple kinematic considerations discussed in Subsection 3.1. The second interaction term on the r.h.s. of Eq.~\eqref{dir} contributes to the process with two
$\hat{\Psi}$-particles in the final state (see the left plot in Fig.~\ref{Fig:fig1}), and is kinematically allowed. We estimate the associated vacuum decay rate
in Subsection 3.1. In particular, we will see that it gives a negligible
contribution to the total decay rate compared to the interaction involving tensor degrees
of freedom.

Schematically, the interaction of the
photons with the tensor modes is given by the Lagrangian,
\begin{equation}
\label{indirph}
{\cal L}_{T-ph} \sim \frac{1}{M_{Pl}} \cdot \hat{h}^{\mu \nu} T^{el}_{\mu \nu} \sim
\frac{1}{M_{Pl}} \hat{h}^{\mu \nu} F_{\mu \alpha} {F^{\alpha}}_{\nu}\; .
\end{equation}
Here $T^{el}_{\mu \nu}$ is the electromagnetic stress-energy tensor; $\hat{h}^{\mu \nu}$ is the canonically normalized field of the helicity-2 graviton
related to the traceless part of the metric, $h^{\mu \nu}$, by
\begin{equation}
\nonumber
\hat{h}^{\mu \nu} \sim M_{Pl} \cdot h^{\mu \nu} \; .
\end{equation}
Scalar-tensor interactions in the
gravity sector are of the form,
\begin{equation}
\label{indirgh0}
{\cal L}_{T-gh} \sim \frac{1}{M_{Pl}} \hat{h}^{\mu \nu} T^{\hat{\Psi}}_{\mu \nu}
\sim \frac{1}{M_{Pl}} \left( \hat{h}^{00} \dot{\hat{\Psi}}^2+c^2_s \hat{h}^{ij} \partial_i
\hat{\Psi} \partial_j \hat{\Psi}  \right) \; ,
\end{equation}
where $T^{\hat{\Psi}}_{\mu \nu}$ is the
stress-energy tensor of the scalar $\hat{\Psi}$. Note that the two terms inside the
parentheses of Eq.~\eqref{indirgh0} are of the same order.
Thus, we can estimate the strength of the scalar-tensor interaction
simply by
\begin{equation}
\label{indirgh}
{\cal L}_{T-gh} \sim \frac{\gamma}{M^3_{Pl}} \cdot \hat{h}^{ij} \partial_i \hat{\Psi} \partial_j \hat{\Psi} \; .
\end{equation}
The corresponding diagram of the vacuum decay into a couple
of photons and ghosts is pictured in Fig.~\ref{Fig:fig1} (right plot),
and contains the propagator of the graviton. We estimate the associated
decay rate in Subsection 3.2.

One comment is in order before we proceed. While we mainly focus on the
electromagnetic interactions of the ghosts, there are a few more, which may trigger potentially
dangerous processes. The first ones involve two
gravitons/neutrinos in the final state. The corresponding interaction Lagrangians
have the same order of magnitude as in the case of photons. However, those processes
leave much weaker signatures in the observational data, and therefore, are practically undetectable. The process with an electron-positron
pair in the final state is irrelevant for another reason. The pair carries the minimal energy $\sim 1~\mbox{MeV}$, which must be balanced
by the outgoing ghosts. We will see, however, that the energies of the particles produced in the
vacuum decay do not exceed $\sim 1~\mbox{keV}$. Hence, this process is forbidden.

\subsection{Vacuum decay into photons from the 4-point contact interaction with ghosts\label{subsection:CI}}
In view of our objectives, it will be enough to perform a rough estimation
of the vacuum decay rate. In particular, we will ignore the interference
between two diagrams in Fig.~\ref{Fig:fig1}, and calculate the
associated rates separately. We start with the case of the $4$-point
contact interaction between photons and ghosts~\eqref{dir}. The corresponding matrix element is estimated by
\begin{equation}
\nonumber
{\cal M}({\bf k}_1, {\bf k}_2; {\bf p}_1, {\bf p}_2) \sim \frac{\gamma}{M^4_{Pl}} \cdot |{\bf k}_1| \cdot
|{\bf k}_2| \; .
\end{equation}
We use the notation ${\bf k}_i$ for the momenta of the photons, and ${\bf p}_i$ for the momenta
of the ghosts. Recall that the coefficient $\frac{\gamma}{M^4_{Pl}}$
originates from the definition~\eqref{candef} of the canonically normalized variable $\hat{\Psi}$;
the factor $|{\bf k}_1|\cdot |{\bf k}_2|$ stems from the derivative
structure of the interactions in Eq.~\eqref{dir}. Modulo the irrelevant phase factors, the decay rate per space-time volume is
given by,
\begin{equation}
\label{rategen}
\Gamma \sim \int \frac{d {\bf p}_1}{|\omega_1|} \frac{d {\bf p}_2}{|\omega_2|}
\frac{d {\bf k}_1}{E_1} \frac{d{\bf k}_2}{E_2} |{\cal M}({\bf k}_1, {\bf k}_2; {\bf p}_1, {\bf p}_2)|^2 \delta^{(4)}(p_1+p_2+k_1+k_2) \; ,
\end{equation}
where $\delta^{(4)}(...)$ is the delta function, which ensures
the conservation of the energy and momentum; $\omega_i \equiv p^{0}_i <0$ and $E_i \equiv k^{0}_i$ denote the
energies of the ghosts and photons, respectively. To handle the integral in Eq.~\eqref{rategen},
it is convenient to introduce the intermediate integrals over fictitious momenta $P$ and $K$~\cite{Sbisa:2014pzo},
\begin{equation}
\int d^4 P \delta^{(4)} (P-p_1-p_2)=1\;, \qquad \int d^4 K \delta^{(4)} (K-k_1-k_2)=1 \; .
\end{equation}
Then, the decay rate can be written
as follows,
\begin{equation}
\label{gammadir}
\Gamma \sim \frac{\gamma^2}{M^8_{Pl}}\int d^4P d^4 K \delta^{4} (K+P) {\cal I}_1 (P)
{\cal I}_2 (K) \; .
\end{equation}
In the case of the direct coupling of the field $\Psi$ to the matter, the integrals ${\cal I}_1 (P)$ and ${\cal I}_2 (K)$ are defined as,
\begin{equation}
\label{i1p}
{\cal I}_1 (P) = \int \frac{d {\bf p}_1}{|\omega_1|} \frac{d {\bf p}_2}{|\omega_2|} \delta^{(4)} (P-p_1-p_2) \;,
\end{equation}
and
\begin{equation}
\label{i2k}
{\cal I}_2 (K)= \int \frac{d {\bf k}_1}{E_1} \frac{d {\bf k}_2}{E_2} |{\bf k}_1|^2 \cdot |{\bf k}_2|^2\delta^{(4)} (K-k_1-k_2) \; .
\end{equation}
These integrals can be evaluated in a straightforward manner. The result for the integral ${\cal I}_1 (P)$ reads
\begin{equation}
\label{i1pest1}
{\cal I}_1 (P)  \sim \frac{M^3_{Pl}}{\gamma^{3/2}}\;,
\end{equation}
(note that it is independent of the momentum $P$). Partially such a huge value of the
integral is explained by the presence of the factors ${|\omega_i |}^{-1}$ in Eq.~\eqref{i1p} (recall
the dispersion relation $|\omega_i| =c_s|{\bf p}_i| \sim \frac{\sqrt{\gamma}}{M_{Pl}}|{\bf p}_i|$). The estimate
for the integral ${\cal I}_2 (K)$ is given by,
\begin{equation}
\label{i2kest1}
{\cal I}_2 (K) \sim K^4_0 \; .
\end{equation}
Substituting Eqs.~\eqref{i1pest1} and~\eqref{i2kest1} into Eq.~\eqref{gammadir} and performing the integration over the fictitious 4-momentum $P$, we get the following
estimate for the decay rate,
\begin{equation}
\nonumber
\Gamma \sim \frac{\sqrt{\gamma}}{M^5_{Pl}} \int d^4 K K^4_0 \; .
\end{equation}
In order to understand the region of integration over the momentum $K$, one
should include kinematic considerations. So, one has
\begin{equation}
\nonumber
|{\bf K}| \lesssim K_0= |{\bf k}_1|+|{\bf k}_2|=
c_s(|{\bf p}_1|+ |{\bf p}_2|) \; .
\end{equation}
Recall now that the spatial momenta of the
ghost particles are bounded from above by the
strong coupling scale, above which the theory is assumed
to be free of instabilities. Hence, $|{\bf K}| \lesssim K_0 \lesssim c_s \Lambda_p$.  As it follows, for $c^2_s \ll 1$, the momenta of the photons are much smaller than those of the ghost particles. Consequently, for the momentum
conservation equation to be obeyed, the outgoing ghost particles must be practically anti-collinear.

Using kinematic considerations, we obtain an order of magnitude expression for the decay rate,
\begin{equation}
\nonumber
\Gamma \sim \frac{\gamma^{9/2} \Lambda^8_p}{M^{13}_{Pl}} \; .
\end{equation}
We see explicitly the huge suppression by a large power of the Planck mass. This fact becomes
particularly prominent upon substituting the estimate for the strong coupling scale, i.e., $\Lambda_p \propto \frac{\gamma^{3/4}}{M^{1/2}_{Pl}}$. We obtain,
\begin{equation}
\label{decsub}
\Gamma \sim \frac{\gamma^{21/2}}{M^{17}_{Pl}} \; .
\end{equation}
By contrasting this decay rate to the observed
flux of photons on the Earth, one can extract a
constraint on the parameter $\sqrt{\gamma}$ and, consequently,
on the strong coupling scale $\Lambda_p$. As we will see in the next Subsection, however,
the decay rate~\eqref{decsub} is sub-dominant compared to that of the process involving
an exchange by the graviton. Hence, the resulting
constraints are expected to be milder.

Here let us pause for an instant to show that the
process with two photons and one ghost in the final state is indeed
forbidden. This follows from kinematic
considerations similar to those discussed above. The momentum ${\bf p}$
of the (only) ghost is defined from the conservation equation, i.e., ${\bf p}=-{\bf k}_1-{\bf k}_2$, where ${\bf k}_1$ and ${\bf k}_2$
are the momenta of the photons. The total energy of the particles
in the final state is given by,
\begin{equation}
\nonumber
|{\bf k}_1|+|{\bf k}_2|-c_s |{\bf k}_1+{\bf k}_2| \gtrsim (|{\bf k}_1|+|{\bf k}_2|) \cdot (1-c_s) >0 \; .
\end{equation}
The inequality on the r.h.s. implies that the energy conservation equation in the
process with one ghost cannot be obeyed. Consequently, this process does not occur.
It is straightforward to generalize the latter statement to any process
of the vacuum decay with one ghost particle in the final state.

\subsection{Vacuum decay into photons mediated by the exchange of a graviton\label{subsection:GE}}

The matrix element for the process involving an exchange by the graviton is
estimated as,
\begin{equation}
\label{matricgrav}
{\cal M}({\bf k}_1, {\bf k}_2; {\bf p}_1, {\bf p}_2) \sim \frac{\gamma}{M^4_{Pl}} \frac{(|{\bf k}_1| \cdot |
{\bf k}_2|) \cdot (|{\bf p}_1| \cdot |{\bf p}_2|)}{q^2} \; .
\end{equation}
Here $(q^2)^{-1}=[(\omega_1+\omega_2)^2 -({\bf p}_1+{\bf p}_2)^2]^{-1}$,---the propagator of the graviton. The decay rate is estimated by the same generic expression~\eqref{rategen}, now with the matrix element~\eqref{matricgrav} substituted in.
Again introducing the integration over the fictitious momenta $P$ and $K$, one can write the
decay rate as in Eq.~\eqref{gammadir} with the integral ${\cal I}_2 (K)$ still given by Eq.~\eqref{i2k},
and the integral ${\cal I}_1 (P)$ defined by,
\begin{equation}
\nonumber
{\cal I}_1 (P) = \frac{1}{P^4}
\int \frac{d {\bf p}_1}{|\omega_1|} \frac{d {\bf p}_2}{|\omega_2|} \cdot |{\bf p}_1|^2\cdot |{\bf p}_2|^2\delta^{(4)} (P-p_1-p_2) \; .
\end{equation}
The factor $1/P^4$ stems from the propagator of the graviton.
To evaluate this integral, it is convenient to make a redefinition of the variable $P_0 =c_s\tilde{P}_0$. Then, the value of the integral can be estimated on the simple dimensional grounds,
\begin{equation}
\nonumber
{\cal I}_1 (P) \sim \frac{1}{P^4} \frac{M^3_{Pl}}{\gamma^{3/2}} \tilde{P}^4_0  \sim \frac{1}{P^4} \frac{M^7_{Pl}}{\gamma^{7/2}}P^4_0 \sim \frac{M^7_{Pl}}{\gamma^{7/2}} \; .
\end{equation}
Note that compared to the case with the direct coupling to the field $\Psi$, we gained the amplification
factor $M^4_{Pl}/\gamma^2$. Therefore, the resulting decay rate is parametrically larger,
\begin{equation}
\nonumber
\Gamma \sim \frac{\gamma^{17/2}}{M^{13}_{Pl}} \; .
\end{equation}
Now, let us contrast our theoretical prediction of the vacuum decay rate to the
experimental data. The number density $n$ of the produced photons is related to the quantity $\Gamma$ by~\cite{Cline:2003gs}
\begin{equation}
\nonumber
n \simeq \Gamma t_0 \; ,
\end{equation}
where $t_0$ denotes the age of the Universe, $t_0 \sim H^{-1}_0$,
and $H_0$ is the current Hubble rate. On the other hand, the measured flux $F$ of the
photons in the range of energies $E_{ph}$ corresponding to the X-rays (keV) and gamma-ray bursts is estimated from~\cite{Hickox:2005dz, Turler:2010pm, Sreekumar:1997un}
\begin{equation}
\nonumber
F \cdot E_{ph} \sim A \cdot \frac{\mbox{keV}}{\mbox{s} \cdot \mbox{cm}^2 \cdot \mbox{sr}} \; .
\end{equation}
We introduced the fictitious dimensionless parameter $A$, which ranges
between $1$ and $100$. The uncertainty here accounts for the slight energy dependence
of the quantity $F \cdot E_{ph}$. We demand that the flux of the photons originating from the
vacuum decay does not exceed the observed one, and that yields the upper bound on the
parameter $\sqrt{\gamma}$,
\begin{equation}
\label{constrgamma}
\sqrt{\gamma} \lesssim 10^{9}~\mbox{GeV} \; ,
\end{equation}
(practically independent of the uncertainty on the parameter $A$). This sets the limit on the strong coupling scale in the theory,
\begin{equation}
\label{constrstrong}
\Lambda_p \lesssim 10~\mbox{TeV} \;,
\end{equation}
(accidentally, it coincides with the scale of experiments at the LHC). The constraint~\eqref{constrstrong}
is seven orders of magnitude less stringent than the limit
of Ref.~\cite{Cline:2003gs} deduced assuming the standard dispersion relation for the ghosts. Furthermore, Eq.~\eqref{constrstrong} demonstrates thirty orders of magnitude
improvement compared to the constraint obtained in the branch of the model plagued by gradient instabilities.

Constraints~\eqref{constrgamma} and~\eqref{constrstrong}
imply that the maximal energies of the produced particles lie in the keV-range.
This follows from Eq.~\eqref{cutoffw},
\begin{equation}
\label{constrstrongw}
\Lambda_{\omega} \lesssim 1~\mbox{keV} \; .
\end{equation}
Therefore, the comparison with the
flux of the cosmic X-rays is justified. The result~\eqref{constrstrongw} has
immediate consequences for the series of processes would be going naively in the presence of the ghosts.
First, it forbids the vacuum decay with an electron-positron pair in the final state. Indeed,
the minimal mass of the pair is of the order~\mbox{MeV} and is parametrically
larger than that allowed by Eq.~\eqref{constrstrongw}. For the similar reasons, the constraint on the
energy of the produced ghosts makes it impossible for lighter particles to decay into heavier ones~\cite{Carroll:2003st}.
So, the hypothetic processes of the electron decay into the muon and two neutrinos, i.e., $e^{-} \rightarrow \mu^{-}+ \nu_e +\bar{\nu}_{\mu}+\mbox{ghosts}$ , or the proton decay with the
neutron, positron and neutrino in the final state, i.e., $p \rightarrow n + e^{+}+\nu_e+\mbox{ghosts}$, do not occur.

In the remainder of the Section, we comment on the alternative ways to constrain
the parameter $\sqrt{\gamma}$ and the strong coupling scale $\Lambda_p$.
First, the vacuum decay triggered by the direct coupling of the ghosts
to the photons is characterized by the smaller rate and, hence, results
into milder limits on those parameters. These read $\sqrt{\gamma} \lesssim 10^{11}~\mbox{GeV}$
and $\Lambda_p \lesssim 10^4~\mbox{TeV}$.

Stronger constraints, of the order of those given in Eqs.~\eqref{constrgamma} and~\eqref{constrstrong},
follow from cosmological considerations. That is, one does not want to overproduce radiation.
Indeed, the energy density of the photons originating from the vacuum decay is estimated by
\begin{equation}
\nonumber
\rho_{ph} \sim c_s \Lambda_p \Gamma t_0 \; .
\end{equation}
The factor $c_s \Lambda_p$ stands for the maximal energy of the
photons, while the factor $\Gamma t_0$ accounts for their number
per unit volume. This should not exceed the total energy density of radiation in the Universe, $\rho_{rad} \sim 10^{-5} H^2_0 M^2_{Pl}$. Substituting the numbers, we again obtain the upper limit~\eqref{constrstrong}
on the strong coupling scale.

Note that the cosmological constraint taken separately could be
essentially relaxed for the following reasons. The energy
of the ghosts produced in the vacuum decay is equal to that
of the photons, but has opposite sign. Furthermore, the condensate of the
ghosts is characterized by a radiation-like equation of state. Hence,
at least naively, the vacuum decay with the photons in the final
state does not affect the cosmological evolution appreciably. This
potentially interesting loophole is irrelevant
for our discussion,---the constraint~\eqref{constrstrong} obtained from the direct
observation of $\mbox{keV}$-photons is strong enough to not worry
about any consequences for cosmology.

The constraint~\eqref{constrgamma} on the parameter $\sqrt{\gamma}$ may have some applications for the
Dark Matter physics (albeit, perhaps, futuristic). Converting it to the upper bound
on the sound speed squared, we have\footnote{The lower limit on the sound speed can be inferred
from Eq.~\eqref{gammalower}. It reads, $c^2_s \gtrsim 10^{-42}$.},
\begin{equation}
\nonumber
c^2_s \lesssim 10^{-20} \; .
\end{equation}
This is 10 orders of magnitude stronger than the limit deduced from the galaxy formation considerations.
Namely, for the bottom-up picture of the large scale structure formation to occur, the constraint $c^2_s \lesssim 10^{-10}$ must be applied~\cite{Capela:2014xta}. The upper bound
here corresponds to the situation, when the formation of dwarf galaxies is suppressed.
Such a large value of the sound speed squared could be relevant for the so called missing satellite problem~\cite{Weinberg:2013aya, Klypin:1999uc, Moore},---the observed number of the dwarf galaxies is much smaller than the one
predicted in the Cold Dark Matter framework. Note, however,
that the mismatch between the observations and the theoretical expectation can be relatively simply explained by the effects of the
baryonic physics~\cite{Sawala:2014xka},---hence, it is not an immediate source of worry.

\section{Discussion}

In the present paper, we showed that the strong coupling
scale of the projectable Ho\v{r}ava--Lifshitz gravity can be raised to
$10$ TeV, upon switching to the ghost unstable branch of the scenario.
This is certainly an advantage over the branch plagued by the gradient instabilities, since now we do not have
the problems with recovering GR at large distances. We reiterate, however, that
the presence of the strong coupling by itself (even a high one) leads to the loss of perturbativity in the Ho\v{r}ava--Lifshitz gravity. In this sense, prospects for renormalizing
the model,---the original motivation behind the Ho\v{r}ava--Lifshitz gravity,---still remain unclear.

Meanwhile, keeping in mind the problems with the UV completion, we can enjoy the rich phenomenology of the
model. Perhaps, the most relevant one is the Dark Matter.
This is quite a generic prediction of the
IR modifications of gravity involving the spontaneous breaking of the
Lorentz symmetry, e.g., the ghost condensate~\cite{ArkaniHamed:2003uy, ArkaniHamed:2005gu}. Not surprisingly, the same observation has been made also in the context
of the projectable Ho\v{r}ava--Lifshitz gravity~\cite{Mukohyama:2009mz, Mukohyama:2009zs}. The specific
feature of this Dark Matter is that the fluid elements always
follow geodesics equation, as it is indicated by the constraint~\eqref{constraint}. This typically implies a
pathology in the model,---attracted by the gravitational force, trajectories corresponding to the
different fluid elements cross at finite times
leading to the so-called caustic singularities\footnote{See Refs.~\cite{Felder:2002sv},~\cite{Babichev:2016hys} for the examples of caustic singularities in the models
with the non-canonical scalar fields.}. On the other hand, that conclusion may alter due to the presence of the higher derivative term as in Eq.~\eqref{IDM},
which may smoother the caustic singularity. The corresponding
mechanism has been discussed in Refs.~\cite{Mukohyama:2009zs, Capela:2014xta} and stems from the
possibility to have regions in space, where gravity acts as a repulsive force (namely,
it turns into anti-gravity). However, numerical simulations capable of verifying or ruling out this mechanism are still pending.

Particle production caused by the vacuum decay opens up an intriguing opportunity
to reheat the Universe even without inflation. Note that
for this scenario to be realized, the parameter $\gamma$ must be sufficiently
large at very early times. Otherwise, one would be able to produce
only very low energetic particles. In fact, the constant $\gamma$ is
implied to follow the renormalization group flow starting from the
values of the order of the Planck mass squared. Apart from the obscure quantum gravity issues,
the time dependence of $\gamma$ appears to be necessary for the production of the
Dark Matter with the correct initial conditions~\cite{Mirzagholi:2014ifa, Ramazanov:2015pha}.

From somewhat more down-to-earth perspective, the production of the keV-photons out of the vacuum
may strongly affect the equilibrium in the early Universe. In particular, the injection of the
out-of-equilibrium photons at the redshifts $z \lesssim 10^{6}$ is expected
to strongly modify the black-body spectrum of the Cosmic Microwave Background~\cite{Chluba:2011hw}\footnote{Although, $keV$ photons are quite separated from the CMB
photons by a large energy gap, things were different at redshifts $z \lesssim 10^6$, where
the CMB distortions are expected to be produced. Namely, in that case, the energy of
CMB photons was in the sub-keV range, thus making them particularly vulnerable
to the energy of non-equilibrium photons.}. It would be certainly interesting to contrast the values of the
$\mu$ and $y$-type distortions following from the scenario at hand with the associated
COBE/FIRAS limits, and furthermore make the predictions for the future PIXIE data. This might be one promising way to improve the constraint on the
parameter $\gamma$. At even higher redshifts, the presence of the keV-photons may have
some impact on the Big Bang nucleosynthesis. Note, however, that the bound energies
of the nuclei are much larger and, thus, do not get destroyed by the injection of the
soft photons. Nevertheless, keV-photons may change the conditions at which
nucleosynthesis proceeds.

\section{Acknowledgments}

We are indebted to E.~Babichev, N.~Bartolo, D.~Blas, M.~Ivanov, P.~Karmakar, M.~Pshirkov and A.~Vikman for many useful comments and discussions. The work of FA is supported by the National Taiwan University (NTU) under Project No. 103R4000 and by the NTU Leung Center for Cosmology and Particle Astrophysics (LeCosPA) under Project No. FI121.

\section*{Appendix A: On the equivalence between the IR limit of projectable Ho\v rava--Lifshitz gravity and mimetic matter scenario.}

At the end of Subsection 2.1, we pointed out that the action for the IR limit of the
projectable Ho\v rava--Lifshitz gravity can be written in the form~\eqref{IDM}. On the other hand, Eq.~\eqref{IDM} describes the
dynamics of the version of the mimetic matter scenario considered in
Refs.~\cite{Chamseddine:2014vna, Capela:2014xta, Mirzagholi:2014ifa, Hammer:2015pcx}. Hence, results of the present paper can
be literally translated into the latter context. The fact that the projectable Ho\v rava--Lifshitz gravity and the mimetic
matter scenario are equivalent was first pointed out in Ref.~\cite{Jacobson:2014mda}\footnote{More precisely, Ref.~\cite{Jacobson:2014mda} links the projectable Ho\v rava--Lifshitz gravity to the version of the scalar Einstein--Aether model considered in Ref.~\cite{Haghani:2014ita}. However, comparing Refs.~\cite{Chamseddine:2014vna, Capela:2014xta, Mirzagholi:2014ifa, Hammer:2015pcx} with
Ref.~\cite{Haghani:2014ita}, one recognizes the latter as a generalization of the mimetic matter scenario.} . The present Appendix serves to prove this statement rigorously.

The proof of the equivalence is based on the following
observation. Let us consider the action of the form,
\begin{equation}
\label{actionsch1}
S= \int {\cal L} (\chi, f)+\lambda (f-F(\chi, \partial_{\mu} \chi, \partial_{\mu \nu} \chi,...)) \; .
\end{equation}
Here $ {\cal L} (\chi, f)$ is the Lagrange function of some variables $\chi$ and $f$ as well as their derivatives, and the Lagrange multiplier $\lambda$ enforces
the constraint $f=F(\chi, \partial_{\mu} \chi, \partial_{\mu \nu} \chi,...)$, so that the field $f$ is a function $F$ of the field $\chi$ and its derivatives.
Then, one can show that the dynamics of the field $\chi$ generated by Eq.~\eqref{actionsch1} is equivalent to that following from the reduced action,
\begin{equation}
\label{actionsch2}
S_r=\int {\cal L} (\chi, f) \left. \right|_{f=F(\chi, \partial_{\mu} \chi, \partial_{\mu \nu} \chi,...)} \; .
\end{equation}
The proof of this statement is given in the Comment 1 of Section~2 of  Ref.~\cite{Pons} (see also the references therein), and we do not repeat
it here. We would like to emphasize that the equivalence between the actions~\eqref{actionsch1} and~\eqref{actionsch2} is not exact, but only with respect to the
dynamics generated for the field $\chi$. Indeed, variation of the
action~\eqref{actionsch1} with respect to the field $f$ gives rise to the equation of motion, which is absent
in the case of the action~\eqref{actionsch2}. This observation will be important for our further discussion.

Now, let us consider the 'true' action for the projectable
Ho\v rava--Lifshitz gravity, i.e., before substituting the
constraint $g^{\mu \nu} \partial_{\mu} \varphi \partial_{\nu} \varphi=1$. Combining Eqs.~\eqref{HLinf},~\eqref{lapsestuck} and~\eqref{K} we get,
\begin{equation}
\label{true}
S_{IR}=S_{GR}+\int d^4 x \sqrt{-g} \left[\frac{\tilde{\Sigma}}{2}\left(g^{\mu \nu} \partial_{\mu} \varphi \partial_{\nu} \varphi-1 \right)+\frac{\gamma}{2} \left(\nabla_{\mu} \left(\frac{\nabla^{\mu} \varphi}{\sqrt{g^{\mu \nu} \partial_{\mu} \varphi \partial_{\nu}\varphi}} \right)\right )^2 \right] \; .
\end{equation}
Here we introduced the notation $\tilde{\Sigma}$ for the Lagrange multiplier field to avoid the confusion in the future. Below we also repeat  the action~\eqref{IDM} of the mimetic matter scenario for the convenience of the references,
\begin{equation}
\label{IDM1}
S_{mim}= S_{GR}+\int d^4 x \sqrt{-g} \left[\frac{\Sigma}{2} \left( g^{\mu \nu} \partial_{\mu} \varphi \partial_{\nu} \varphi-1 \right)+\frac{\gamma}{2} (\square \varphi )^2 \right] \; .
\end{equation}
The point is to show that both actions~\eqref{true} and~\eqref{IDM1}  are equivalent to the following one,
\begin{equation}
\label{formal}
S= S_{GR}+\int d^4 x \sqrt{-g} \left[\frac{\Sigma}{2} \left( g^{\mu \nu} \partial_{\mu} \varphi \partial_{\nu} \varphi-X \right)+\frac{\tilde{\Sigma}}{2} \left(X-1\right)+\frac{\gamma}{2} \left(\nabla_{\mu} \left(\frac{\nabla^{\mu} \varphi}{\sqrt{X}} \right)\right )^2 \right]
\; ,
\end{equation}
and, therefore, they are equivalent between each other. In Eq.~\eqref{formal}, $X$  is the new variable, which obeys the constraint $X=1$ enforced by the Lagrange multiplier field $\tilde{\Sigma}$. First, it is obvious that
 the actions~\eqref{formal} and~\eqref{IDM1} match the generic ones~\eqref{actionsch1} and~\eqref{actionsch2},  respectively,
with the field  $\lambda$ in Eq.~\eqref{actionsch1} understood as the Lagrange multiplier $\tilde{\Sigma}$ in Eq.~\eqref{formal}. Hence, Eqs.~\eqref{formal} and~\eqref{IDM1} describe the same dynamics with respect
to the fields $\Sigma$, $\varphi$ and the metric $g^{\mu \nu}$.
On the other hand, the actions~\eqref{formal} and~\eqref{true} are analogous to those of Eqs.~\eqref{actionsch1} and~\eqref{actionsch2}, respectively, with the field $\lambda$ in Eq.~\eqref{actionsch1}  understood as the Lagrange multiplier $\Sigma$ in Eq.~\eqref{formal}, and the associated constraint given by
$X=g^{\mu \nu} \partial_{\mu} \varphi \partial_{\nu} \varphi $. Hence,  the actions~\eqref{formal} and~\eqref{IDM1} describe the same dynamics with respect
to the fields $\tilde{\Sigma}$, $\varphi$ and the metric $g^{\mu  \nu}$. To summarize, the actions Eqs.~\eqref{true},~\eqref{IDM1},~\eqref{formal}, and, consequently, the IR limit of the projectable Ho\v rava--Lifshitz gravity and the mimetic
matter scenario, are equivalent.

Let us make one important observation here. Strictly speaking, the discussion above implies only the equivalence in the dynamics of the metric $g^{\mu \nu}$ as well as the khronon
field $\varphi$. Namely, provided that they start from  the same initial conditions in both models, they follow the same evolution at later times.
However, this is not true for the fields $\Sigma$ and $\tilde{\Sigma}$  (that is why we chose the
different notations for them)\footnote{As the attentive reader could notice, the reason is that the  projectable Ho\v rava--Lifshitz model is equivalent to the model~\eqref{formal} with respect to the set of the
fields $(\tilde{\Sigma}, \varphi, \Psi,...)$. At the same time, the mimetic matter scenario is  equivalent to the model~\eqref{formal} with respect to the set of the fields $(\Sigma, \varphi, \Psi,...)$.}.  To understand the difference in treating the Lagrange multiplier fields  in two models, it is enough to vary the action~\eqref{formal} with respect to the field $X$, and then set the latter to unity
in the end. This gives,
\begin{equation}
\label{Lagrange}
\Sigma -\tilde{\Sigma}= \gamma \nabla_{\rho} \square \varphi  \nabla^{\rho} \varphi \; .
\end{equation}
Therefore, one should be cautious, when keeping the fields $\Sigma$ and $\tilde{\Sigma}$ as the independent variables. In particular, setting them to be equal at the initial Cauchy surface will result into different dynamics in the two models. On the other hand, with the choice of the
independent variables $(\varphi, g^{\mu \nu})$\footnote{This has been indeed our strategy in Subsection 2.2. That is, the Lagrange multiplier field is not
present in the quadratic action~\eqref{action}: it drops off upon substituting the constraint~\eqref{linc}.}, the dynamics will be the same, i.e., the predictions of the two models will be physically
indistinguishable.  The technical reason is that  the shift~\eqref{Lagrange} is exactly compensated by the associated shift of the stress-energy tensor calculated in the IR
limit of the projectable Ho\v rava--Lifshitz gravity $T^{IR}_{\mu \nu}$ compared to that of the mimetic matter scenario $T^{mim}_{\mu \nu}$. Namely, these are related to each
other by,
\begin{equation}
\label{set}
T^{IR}_{\mu \nu}=T^{mim}_{\mu \nu} (\Sigma \rightarrow \tilde{\Sigma})+\gamma \nabla_{\rho} \varphi \nabla^{\rho} \square \varphi \nabla_{\mu} \varphi \nabla_{\nu} \varphi \; ,
\end{equation}
 where the tensor $T^{mim}_{\mu \nu}$  is given by~\cite{Chamseddine:2014vna},
\begin{equation}
\label{mimstress}
T^{mim}_{\mu \nu}=\Sigma \nabla_{\mu} \varphi \nabla_{\nu} \varphi +\gamma \left(\nabla_{\mu} \varphi \nabla_{\mu} \square \varphi+\frac{1}{2} (\square \varphi)^2 \right) g_{\mu \nu}-\gamma (\nabla_{\nu} \varphi
\nabla_{\mu} \square \varphi +\nabla_{\nu} \square \varphi \nabla_{\mu} \varphi) \; .
\end{equation}
Combining Eqs.~\eqref{Lagrange},~\eqref{set} and~\eqref{mimstress}, we get $T^{IR}_{\mu \nu}-T^{mim}_{\mu \nu}=(\tilde{\Sigma}-\Sigma+\gamma \nabla_{\rho} \varphi \nabla^{\rho} \square \varphi )\nabla_{\mu} \varphi \nabla_{\nu} \varphi=0$.
This completes the proof.

\section*{Appendix B: Curing ghost instabilities beyond projectable Ho\v{r}ava--Lifshitz gravity}
Here, we briefly discuss one possible way to simultaneously
cure ghost instabilities and retain the model perturbative. For this purpose,
we include the operator~\eqref{Stuck} into the
analysis. That operator, we remind, has a dimension higher than
marginal. Therefore, it was not considered in the bulk
of the paper. In the unitary gauge, the quadratic action for an extended model including the UV operators~\eqref{LVHL} is given by
\begin{equation}
\label{actionapp}
\begin{split}
\delta_2 S &= \frac{1}{16\pi G} \int d^4x \Bigl[-6 \dot{\Psi}^2-2\Psi\Delta \Psi +4\dot{\Psi}(\Delta \dot{E}-\Delta B) + \frac{\alpha}{M^2_*}\Psi \Delta^2 \Psi+...\Bigr] + \\
&+\frac{\gamma}{2}\int d^4x  \cdot \Bigl[3 \dot{\Psi}+\Delta B-\Delta \dot{E} \Bigr] \hat{A} \left[3 \dot{\Psi}+\Delta B-\Delta \dot{E}\right] \;  ,
\end{split}
\end{equation}
where we assume the Minkowski background. Here $\alpha$ is an order one constant governing the relevant operator and the ellipsis
stand for the contributions from the marginal operators. We introduced the operator $\hat{A}$ defined by,
\begin{equation}
\nonumber
\hat{A} \equiv \Bigl(1-\frac{\beta M^2_{Pl}}{\gamma M^2_*} \Delta \Bigr) \; ,
\end{equation}
where $\beta$ is an order one constant governing the operator~\eqref{Stuck}.

Varying the action~\eqref{actionapp} with respect to the combination $B-\dot{E}$, one gets
the constraint equation,
\begin{equation}
\Delta B-\Delta \dot{E}=\frac{\dot{\Psi}}{4\pi \gamma G \hat{A}}-3\dot{\Psi} \; .
\end{equation}
Substituting this back into the action~\eqref{actionapp}, one gets
\begin{equation}
\nonumber
\delta_2 S =\int d^4x \left(6 \dot{\Psi}^2-\frac{\dot{\Psi}^2}{2\pi \gamma G \hat{A}}-2\Psi \Delta \Psi
+\frac{\alpha}{M^2_*}\Psi \Delta^2 \Psi  \right) \; .
\end{equation}
Switching to the Fourier space analysis, we observe that for small momenta $|{\bf p}| \lesssim M_*$, the
second term on the r.h.s. dominates over the first one. Hence, the
kinetic term has a ghost-like sign in this regime. On the other hand,
for larger momenta $|{\bf p}| \gtrsim M_*$, the first term is dominant,
and we recover the positive sign of the kinetic term.

This solution is not without problems, though. Indeed, that way of curing ghost instabilities is
at risk of getting gradient instabilities instead. The problem can be avoided by a proper tuning of the
constants $\alpha$ and $\beta$.

\end{document}